\documentclass{aa}  

\usepackage{graphicx}
\usepackage{txfonts}
\usepackage{lipsum}
\usepackage{subcaption}         
\usepackage{lscape}             
\usepackage{placeins}           
\usepackage{multirow}
\usepackage{longtable}
\usepackage{tabularx}
\usepackage{booktabs}
\usepackage[colorlinks=true,linkcolor=blue,citecolor=blue, urlcolor=blue]{hyperref}

\begin{document}

\title{Chemo-dynamical reconstruction of Milky Way globular cluster progenitors: Age--metallicity relations and the universality of multiple stellar populations}

\titlerunning{MW assembly history and the universality of MPs}
\subtitle{}

   \author{Carmela Lardo\inst{1,2}\fnmsep\thanks{carmela.lardo2@unibo.it}
        \and David Valcin\inst{3}
        \and Raul Jimenez\inst{4,5}
        }

   \institute{Dipartimento di Fisica e Astronomia ``Augusto Righi''–Universit\`a di Bologna, via Piero Gobetti 93/2, I-40129 Bologna, Italy.
   \and INAF - Osservatorio di Astrofisica e Scienza dello Spazio di Bologna, via Piero Gobetti 93/3, I-40129 Bologna, Italy.
   \and Berkeley Center for Cosmological Physics and Department of Physics, University of California, Berkeley, CA 94720, USA.
   \and ICC, University of Barcelona, Mart\' i i Franqu\` es, 1, E08028
Barcelona, Spain.
    \and ICREA, Pg. Lluis Companys 23, Barcelona, 08010, Spain.}

   \date{Received 12 March 2026/ Accepted 17 April 2026}

\abstract
{Globular clusters encode the hierarchical assembly history of the Milky Way
and the internal physics of multiple stellar populations. Reconstructing
progenitor-specific age--metallicity relations requires stellar parameters
free from helium-driven age biases, yet whether multiple-population properties
carry an environmental imprint remains an open question.}
{We reconstructed progenitor-specific age--metallicity relations for 69 Galactic
globular clusters using homogeneous stellar parameters derived while modelling
multiple stellar populations, and tested whether helium-related multiple-population
properties depend on progenitor origin once cluster mass and metallicity
are controlled for.
}
{Ages, helium spreads ($\delta Y$), mean helium abundances ($\bar{Y}$), and
first-population fractions ($f_{\rm P1}$) were drawn from hierarchical Bayesian
colour--magnitude diagram modelling. Progenitor families were identified via
probabilistic chemo-dynamical clustering, age--metallicity relations
reconstructed within a hierarchical Bayesian framework, and
multiple-population indicators tested for environmental dependence
using regression models with sensitivity tests.}
{Enrichment timescales are broadly consistent with $\tau \lesssim 2$\,Gyr,
though individual progenitors prefer shorter values when fitted independently
and inter-progenitor differences cannot be resolved at the present sample size.
The primary distinction is the extent of chemical evolution: most systems reach
$\Delta[\rm Fe/H] \simeq 1.1$--$1.3$\,dex, while Sagittarius achieves
$\Delta[\rm Fe/H] \simeq 1.6$\,dex and higher terminal metallicities.
Progenitor masses identify Gaia--Sausage--Enceladus and low-energy group Nas the dominant events. 
Neither $\delta Y$ nor
$\bar{Y}$ shows a significant progenitor dependence, and the mass
multiple-population scaling is indistinguishable across in situ and accreted
systems; Sequoia clusters alone show higher $f_{\rm P1}$ at fixed mass
and metallicity.}
{Age--metallicity relations carry fossil signatures of the chemical evolution
and mass hierarchy of progenitor galaxies. The extent of enrichment, but not
its pace, distinguishes progenitor systems. Helium enrichment amplitude is
primarily regulated by cluster mass and blind to environment, pointing to
universal cluster-scale formation physics. The sole robust exception of
a residual progenitor dependence is in $f_{\rm P1}$, suggesting the enriched-star
fraction retains a secondary environmental imprint.}


   \maketitle

\nolinenumbers

\section{Introduction}

Globular clusters (GCs) occupy a unique position among tracers of galaxy assembly: they form predominantly at high redshifts \citep{GC97, Trenti}, survive for a Hubble time, and carry in their stellar populations both the chemical imprint of their birth environments and the dynamical memory of the galaxies in which they formed \citep[e.g.][]{kruijssen19, kruijssen20, forbes20}. Within the $\Lambda$ cold dark matter ($\Lambda$CDM) paradigm, the Milky Way (MW) assembled hierarchically through a sequence of mergers and accretion events in which smaller systems were incorporated into progressively larger dark matter haloes \citep{WhiteRees78}. The MW's GC system preserves a fossil record of this process that is, in principle, legible through the kinematics, ages, and chemical abundances of individual clusters.

In the \textit{Gaia} era 
\citep{gaia18,gaia23}, integrals-of-motion analyses have established that 
the Galactic GC system is composite: roughly $40$--$45\%$ of clusters 
were likely accreted from now-disrupted satellite galaxies, with the 
remainder formed in situ \citep{massari19,massari23, naidu20, naidu21,
youakim23,youakim25,chen24}. The major accreted components — Gaia--Enceladus--Sausage 
(GSE; \citealp{helmi18,belokurov18}), Sagittarius \citep[Sgr;][]{ibata94,law10,
ruizlara20}, Sequoia \citep[Seq;][]{barba19,myeong19}, the Helmi streams 
\citep[H99;][]{helmi99,koppelman19b}, and proposed low-energy inner structures 
\citep{massari19,pfeffer20,horta21,malhan22,massari26} — together encode 
the hierarchical merger history of the Galaxy.

Kinematics alone, however, do not uniquely determine cluster origins. 
Dynamically heated in situ clusters can overlap with accreted 
debris in integrals-of-motion space, particularly at low orbital energies. 
The age--metallicity relation (AMR) provides an independent and physically 
motivated diagnostic: because galaxies of different masses enrich at 
different rates, their GC systems trace distinct loci in age--metallicity 
space \citep{brodie06,leaman13,massari19,forbes20,callingham22}. Combining kinematics 
with AMRs therefore offers a more robust basis for reconstructing the 
assembly history than either approach alone.

Realising this potential requires stellar population parameters that are 
both homogeneous and internally consistent. At old ages, systematic age 
uncertainties of a few hundred megayears are comparable to the temporal separation 
between early merger episodes \citep{wagner17}. A particularly important source of bias is helium: an enhanced helium abundance shifts the isochrone morphology in ways that can mimic younger ages if not modelled explicitly 
\citep{salaris05,dotter10,valcarce12}. This matters because virtually all 
GCs host multiple stellar populations (MPs) — characterised by 
light-element abundance variations and, in many clusters, measurable helium 
spreads \citep{gratton04,gratton12,gratton19,bastian18,milone18,milone22} 
— and the MP fractions and helium enrichment vary systematically between 
clusters. AMR studies that ignore MPs therefore risk propagating 
helium-driven age biases into reconstructions of Galactic assembly 
\citep{milone18,cabrera22}.

Whether MPs themselves carry information about the environment in which 
clusters formed is a separate and largely open question. MP properties — 
helium spreads, enriched population fractions, and light-element variation 
amplitudes — correlate strongly with cluster mass \citep{milone17,
bastian18}, but it remains unclear whether they also depend on the 
host-galaxy environment. The most systematic prior test — comparing chromosome-map properties \citep{milone17} across Magellanic Cloud and Galactic GCs — found no significant environmental dependence beyond the mass scaling, though that analysis employed a coarse in situ--accreted distinction without MP-aware stellar parameters \citep{milone20}. Whether finer progenitor resolution combined with helium-based indicators would reveal a residual environmental signal has not been tested. 
If indeed MPs correlate with environment, this would constitute an additional 
fossil record of galaxy assembly; if they do not, their formation is 
governed by universal cluster-scale physics independent of the large-scale 
galactic context. Distinguishing these scenarios requires a sample of 
clusters with homogeneous MP parameters spanning multiple progenitor 
systems — a dataset that had until now been unavailable.

Here we present a chemo-dynamical reconstruction of the MW GC system that 
addresses both issues simultaneously. 
We combined homogeneous stellar 
population parameters inferred while explicitly modelling MPs 
\citep{valcin21,valcin25,valcin26} with a probabilistic dynamical 
classification of cluster progenitors, and used the resulting dataset to 
(i) reconstruct progenitor-specific AMRs within a hierarchical Bayesian 
framework and (ii) test whether MP properties — helium spreads ($\delta Y$), 
mean helium abundances ($\bar{Y}$), and first-population fractions 
($f_{\rm P1}$) — follow universal scaling relations with cluster mass or 
retain a measurable imprint of the galactic environment in which clusters 
formed.

All previous studies that tested for an environmental dependence of MP properties employed coarse in situ--accreted binary classifications \citep{milone20}, used pseudo-colour proxies sensitive to a combination of abundance variations rather than to helium specifically \citep{milone17}, or relied on stellar parameters derived under the single-population assumption and were therefore subject to systematic helium-driven age biases~\citep{salaris05}. No study has yet combined MP-aware stellar parameters, a multi-progenitor chemo-dynamical classification, and helium-specific indicators within a single homogeneous sample — a combination that is essential to disentangle the environmental from the cluster-mass signal.

The paper is organised as follows. Section~\ref{sec:data} describes the 
data. Section~\ref{sec:methods} presents the chemo-dynamical clustering, 
hierarchical AMR modelling, and MP statistical framework. 
Section~\ref{sec:results} reports the results, and 
Sect. \ref{sec:discussion} discusses the implications for MW assembly 
and the universality of MP formation.

\section{Data}\label{sec:data}

\subsection{Stellar population parameters}

Our sample comprises 69 Galactic GCs with high-quality \textit{Hubble} Space Telescope (HST) multi-band 
photometry spanning a broad range in metallicity, mass, and dynamical 
properties. Cluster ages, $[\rm Fe/H]$, mean helium abundances ($\bar{Y}$), 
helium spreads ($\delta Y$), and first-population fractions ($f_{\rm P1}$) are 
adopted from \citet{valcin26}. These are inferred via hierarchical Bayesian 
modelling of the full colour--magnitude diagram (CMD) morphology, explicitly accounting for two coexisting 
stellar populations sharing a common age and metallicity but differing in 
helium abundance and relative fraction. Helium is parametrised through 
$\bar{Y}$ and the inter-population difference ($\delta Y$), a formulation that 
reduces the degeneracies arising when absolute helium abundances are sampled 
directly.

Because age and helium are partially degenerate in CMD fitting, age posteriors 
are often non-Gaussian. \citet{valcin26} represent them as Gaussian mixture 
models, preserving asymmetric uncertainties and extended tails; these 
Gaussian mixture 
model representations are propagated directly into our hierarchical AMR modelling. 
Typical $68\%$ credible uncertainties are $\approx 1.1$\,Gyr in age, 
$\approx 0.15$\,dex in $[\rm Fe/H]$, and $\approx 0.03$ in $\bar{Y}$, with 
modest cluster-to-cluster variation (0.9--1.7\,Gyr, 0.14--0.17\,dex, and 
0.028--0.033, respectively), indicating uniform precision across the sample. 
Ages of the oldest components are statistically consistent with classical 
single-population determinations, confirming that explicitly modelling MPs 
improves internal consistency without shifting the absolute age scale. 
Extensive internal validation tests are presented in \citet{valcin26}.

\begin{figure*}
\centering
\includegraphics[width=\textwidth]{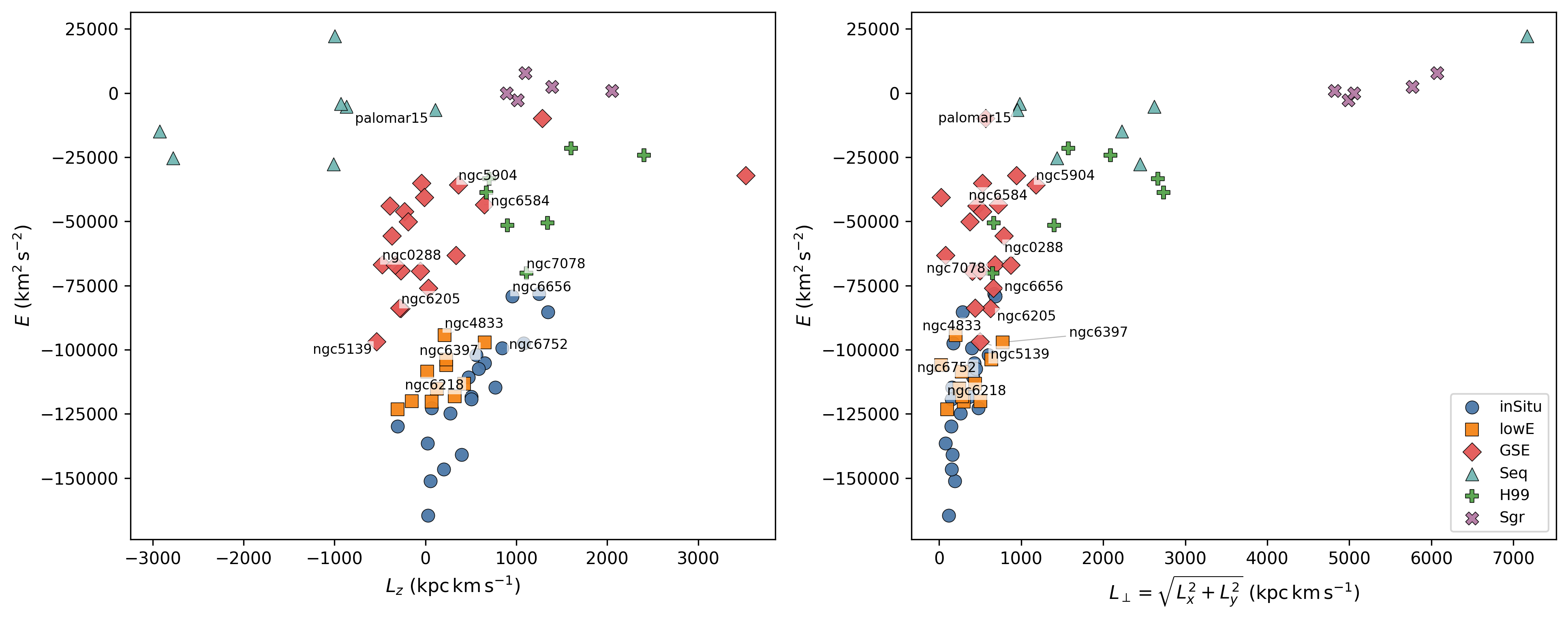}
\caption{Integrals-of-motion projections for the MW GC sample coloured by our final six-family probabilistic classification. 
    {\em Left:} $E$--$L_z$ plane. {\em Right:} $E$--$L_\perp$. 
    Colours and markers denote the assigned family (see the legend). We define the classification confidence using the maximum posterior probability ($p_{\max}$) and the probability margin ($\Delta p = p_{\max}-p_{\rm 2nd}$) between the two most probable classes. Objects are considered secure when $p_{\max}\ge0.80$ and $\Delta p\ge0.20$, likely when $p_{\max}\ge0.60$ and $\Delta p\ge0.10$, and ambiguous otherwise. For clarity, cluster names are shown only for the objects with likely or ambiguous classifications (see Table~\ref{tab:cluster_summary}).}
\label{fig:ELz_ELperp_final}
\end{figure*}

\subsection{Structural and dynamical parameters}

The stellar population parameters described above are complemented by structural and orbital quantities, which serve as both as dynamical classification inputs and as control variables in the MP analysis.

Orbital parameters are computed with \texttt{galpy} \citep{bovy15} via a
Monte Carlo (MC) framework ($N_{\rm MC} = 500$ realisations per cluster), using
proper motions and covariances from \citet{vasiliev21}. Heliocentric distances
and line-of-sight velocities are sampled from Gaussians centred on their
measured values; proper motions are drawn from a bivariate normal
incorporating the published $\mu_{\alpha*}$--$\mu_\delta$ covariance. 
Orbits are integrated backwards for 5\,Gyr, sufficient to sample multiple orbital periods for all clusters and robustly estimate orbital extrema, in the \texttt{MWPotential2014} potential \citep{bovy15},
with the halo normalisation modestly increased relative to the default
to bring the implied MW halo mass closer to recent literature estimates, 
augmented by a rotating Dehnen bar \citep{dehnen00} with fixed pattern 
speed $\Omega_{\rm bar}$, ensuring that the Jacobi energy, 
$E_J = E - \Omega_{\rm bar}L_z$, is conserved in the rotating frame and 
constitutes a well-defined dynamical label for classification.

From each orbit we extracted the $r_{\rm peri}$, $r_{\rm apo}$, eccentricity ($e$), $z_{\rm max}$, $L_z$, $|{\bf L}|$, actions $(J_r, J_z)$, the total specific energy ($E$), and the Jacobi energy ($E_J$). Actions were computed via the St\"{a}ckel approximation \citep{sanders15},
which provides accurate estimates across a wide range of orbit types in
realistic axisymmetric potentials; residual errors are largest for highly radial, resonant, or chaotic orbits, which represent a small minority of the
present sample.
All quantities are summarised by their 16th, 50th,
and 84th MC percentiles.

Structural parameters  -- half-light radius ($r_{\rm h}$), core radius ($r_{\rm c}$),
mass ($M$), tidal radius ($r_{\rm t}$), central density ($\rho_{\rm c}$), and
central velocity dispersion ($\sigma$) -- are taken from the \citet{baumgardt18}
catalogue, derived from dynamical modelling of HST and \textit{Gaia}
kinematics. Cluster mass is used as a control variable throughout the MP
analysis, given its well-established influence on helium spreads
\citep{milone17}.

\section{Methods}\label{sec:methods}

The analysis proceeds in three connected stages. We first assigned each cluster a probabilistic progenitor family using chemo-dynamical clustering (Sect. \ref{sec:clustering}), then reconstructed progenitor-specific enrichment histories within a hierarchical Bayesian framework (Sect. \ref{sec:hierarc_modelling}), and finally tested whether multiple-population properties depend on the inferred environment after controlling for cluster mass and metallicity (Sect. \ref{sec:MP_methods_MPs}).

\subsection{Chemo-dynamical clustering}\label{sec:clustering}

We constructed a probabilistic classification of Galactic GCs using five
features per cluster: the Jacobi energy in the rotating bar frame ($E_{J}$), the vertical angular momentum ($L_z$), the perpendicular angular
momentum, $L_\perp = \sqrt{L_x^2 + L_y^2}$ (a proxy for orbital
inclination), the radial action ($J_r$), and the mean metallicity
($[\mathrm{Fe/H}]$), all evaluated at their median MC values.
These features span the energy, angular momentum, and radial action
dimensions of orbital phase space; metallicity provides an independent
chemical dimension that breaks the energy--$L_z$ degeneracy between
in situ and accreted clusters at low binding energy

Six dynamical families were considered: GSE, 
Sgr, Seq, in situ, low-energy (lowE), and 
H99. For each family we fitted a multivariate Gaussian in 
standardised feature space using literature anchor clusters with robust prior 
associations \citep{massari19,massari23,youakim25}. To mitigate over-fitting 
and allow physically motivated overlap between families, we applied strong 
covariance regularisation and softened class priors proportional to 
$\sqrt{N_{\rm anchor}}$.

Posterior membership probabilities were computed via Bayesian inference
from the class likelihoods and priors. All clusters receive a family
assignment to their highest-probability class; classification confidence
is communicated through `secure', `likely', and `ambiguous' tiers rather than a hard probability threshold
(Appendix~\ref{sec:appendix_classification}). We verified that
replacing hard assignments with probability-weighted regression leaves
all MP results unchanged, as expected given that the majority of
clusters have $p_{\rm max} \geq 0.90$.

\subsection{Hierarchical modelling of the AMR}\label{sec:hierarc_modelling}

We modelled the AMRs within a hierarchical Bayesian framework that accounts for
the small number of clusters per progenitor. Rather than fitting each
progenitor independently, we adopted a partially pooled model in which enrichment
parameters vary between progenitors but are regularised by a common
population-level distribution, stabilising inference for sparsely sampled
systems while permitting genuine inter-progenitor differences.

Each progenitor AMR is described by a truncated saturating exponential in
look-back time ($t$),
\begin{equation*}
    [\mathrm{Fe/H}](t) \;=\; [\mathrm{Fe/H}]_0 \;+\;
    \Delta[\mathrm{Fe/H}]\left(1 - e^{-(L_0 - t)/\tau}\right),
    ~L_{\rm stop} \leq t \leq L_0,
\end{equation*}
where $L_0$ is the onset look-back time, $\tau$ is the enrichment timescale
controlling the curvature of the rise, $\Delta[\mathrm{Fe/H}]$ is the total
metallicity increase, and $[\mathrm{Fe/H}]_{\rm stop} \equiv
[\mathrm{Fe/H}]_0 + \Delta[\mathrm{Fe/H}]$ is the terminal metallicity at
which enrichment is truncated. For $t < L_{\rm stop}$ the metallicity is held
fixed at $[\mathrm{Fe/H}]_{\rm stop}$, encoding the cessation of star
formation upon accretion into the MW potential. This parametrisation
separates the pace of enrichment ($\tau$, shared across progenitors
to within population-level scatter) from its extent
($\Delta[\mathrm{Fe/H}]$, which varies between systems), a distinction that
proves central to the results of Sect. \ref{sec:amr}.

Asymmetric age uncertainties from the CMD analysis are propagated via latent
true ages with split-normal likelihoods. Convergence was verified with the
Gelman--Rubin diagnostic ($\hat{R} \leq 1.003$ for all parameters) and
confirmed by effective sample sizes exceeding $10^3$, indicating that the
indicating that the chains are well mixed and that posterior summaries -- including credible intervals for $\tau$, 
$\Delta[\rm Fe/H]$, and $[\rm Fe/H]_{\rm stop}$ -- are numerically stable.

\subsection{Statistical analysis of multiple-population 
indicators}\label{sec:MP_methods_MPs}

To test whether MP properties depend on progenitor environment, we fitted linear 
models of the form
\begin{equation*}
Y = \beta_0 + \beta_1 \log M + \beta_2 [\mathrm{Fe/H}] + \epsilon
\end{equation*}
for each MP indicator $Y \in \{\delta Y,\, f_{\rm P1},\, \bar{Y}\}$, with 
standardised predictors to facilitate comparison of effect sizes. Progenitor 
family is then added as a categorical variable, and its significance assessed 
via Type-III analysis of variance (ANOVA)  with HC3 robust standard errors to account for 
heteroscedasticity\footnote{We verified that replacing hard family assignments with probability-weighted regression — where each cluster contributes to each family in proportion to its posterior membership probability — leaves all results unchanged. This is expected given that the majority of clusters have $p_{\rm max} \geq 0.90$ (Table~\ref{tab:cluster_summary}), so the two approaches are numerically near-identical for the present sample.}. The interaction terms
\begin{equation*}
Y \sim \log M + [\mathrm{Fe/H}] + \mathrm{prog}
+ (\log M \times \mathrm{prog})
+ ([\mathrm{Fe/H}] \times \mathrm{prog})
\end{equation*}
test whether the mass--MP or metallicity--MP slopes themselves differ between 
progenitors; a null result implies universal scaling relations across in situ 
and accreted systems. Residual MP indicators, computed after subtracting the 
best-fitting mass and metallicity dependence, are used to visualise 
environmental offsets directly.

To assess the robustness of any detected progenitor signal, we repeated the 
regression after removing borderline Sequoia members individually and in 
combination, and performed a permutation test ($n = 10{,}000$ shuffles) to 
obtain an empirical null distribution for the progenitor $F$-statistic 
independent of parametric assumptions. We applied an analogous framework to test for dynamical evolution effects, 
examining correlations between MP indicators and orbital parameters 
($r_{\rm peri}$, $e$, $z_{\rm max}$, $t_{\rm rh}$) and mass-loss proxies 
derived from initial cluster masses \citep{baumgardt18}.

   \begin{figure}
        \centering
        \includegraphics[width=0.9\hsize]{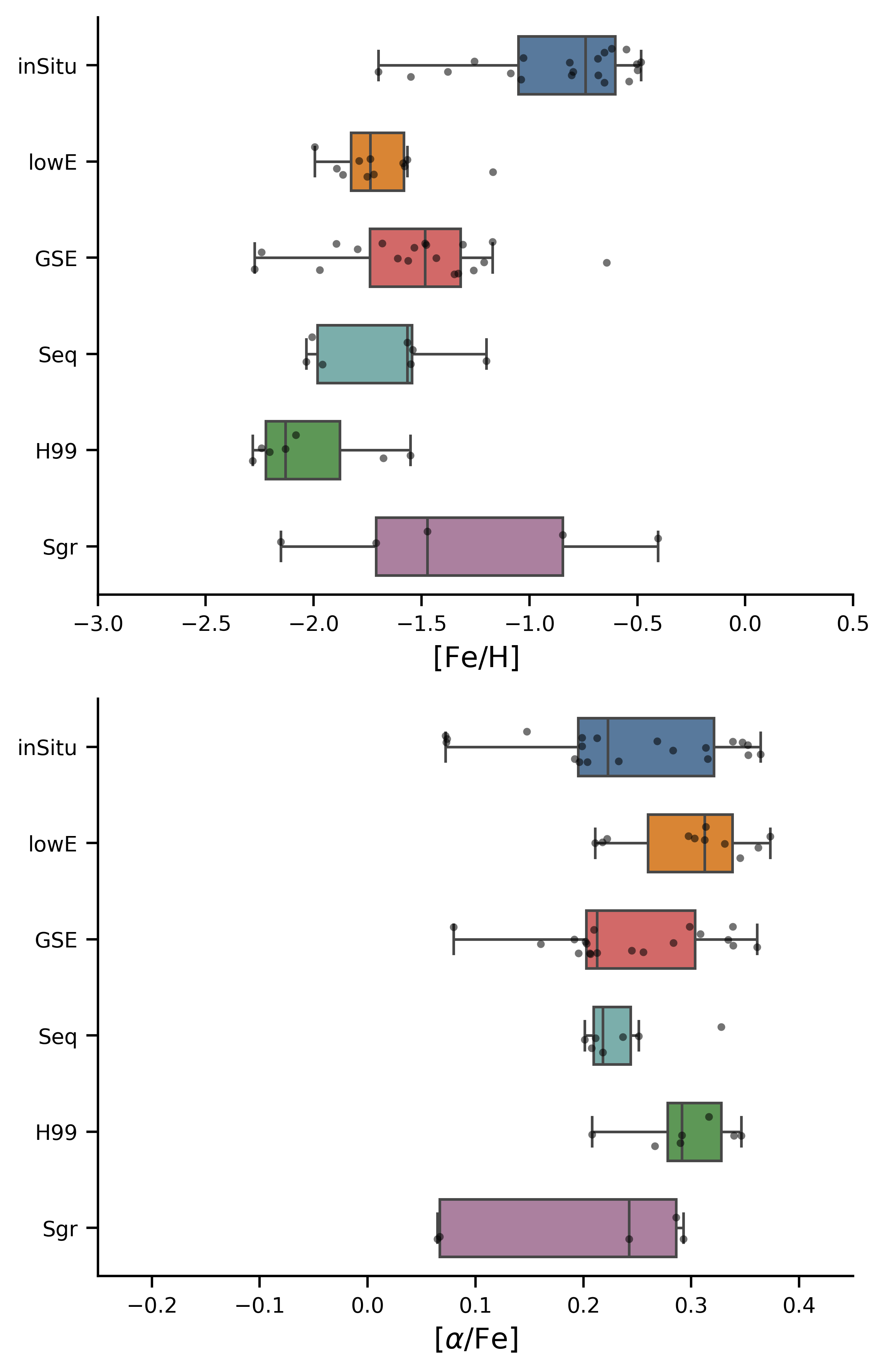}
        \caption{Box-and-whisker plots showing the distributions of $[\mathrm{Fe/H}]$ (\textit{top}) and $[\alpha/\mathrm{Fe}]$ (\textit{bottom}) for the GCs associated with each progenitor family.}
        \label{Fig:kernel}%
    \end{figure}

\section{Results}\label{sec:results}

\subsection{Dynamical families and chemical properties}
\label{sec:dyn_chem}

Figure~\ref{fig:ELz_ELperp_final} shows the integrals-of-motion projections for the MW GC sample coloured by the six-family probabilistic classification (Sect. \ref{sec:clustering}; Appendix~\ref{sec:appendix_classification}). The major accretion structures recover their expected dynamical signatures. Sagittarius occupies a compact, isolated locus in action--energy space, consistent with debris from a comparatively recent and not-yet fully phase-mixed accretion event \citep{ibata94,law10,ruizlara20}. Sequoia clusters lie predominantly on retrograde orbits (negative $L_z$), consistent with their identification as the debris of a retrograde merger \citep{myeong19,barba19}. The GSE family populates a broad distribution centred near $L_z \approx 0$, as expected for the debris of a highly radial merger \citep{belokurov18,helmi18}. The Helmi-stream and low-energy families partially overlap with GSE in the central action--energy region, consistent with the strong phase mixing expected for early accretion debris in the inner halo \citep{helmi99,koppelman19b,massari19,pfeffer20,horta21,massari26}.

The [Fe/H] distributions across the six families 
(Fig.~\ref{Fig:kernel}, top) broadly follow the stellar mass--metallicity 
relation: the in situ population is the most metal-rich 
($\langle[\rm Fe/H]\rangle = -0.77 \pm 0.05$\,dex), while accreted systems 
span progressively lower metallicities ordered roughly by progenitor mass. 
H99 is the most metal-poor ($-2.11 \pm 0.09$\,dex), followed by Sequoia 
($-1.68 \pm 0.11$\,dex), lowE ($-1.72 \pm 0.07$\,dex), GSE 
($-1.50 \pm 0.06$\,dex), and Sagittarius ($-1.46 \pm 0.15$\,dex). 
An ANOVA test confirms that these differences are highly significant 
($p = 1.2 \times 10^{-9}$), with the in situ population more 
metal-rich than GSE, Sequoia, and lowE at $p < 10^{-3}$, and H99 
significantly more metal-poor than both GSE ($p \simeq 0.049$) and 
Sagittarius ($p \simeq 0.023$). 

Most accreted families show modest spreads ($\sigma_{\rm int} \simeq 0.16$--$0.34$\,dex), 
consistent with regulated pre-accretion enrichment. Sagittarius is the 
exception with $\sigma_{\rm int} = 0.53^{+0.14}_{-0.17}$\,dex, suggestive 
of prolonged or multi-episode enrichment before and during disruption. The 
in situ family also shows elevated scatter 
($\sigma_{\rm int} = 0.31^{+0.06}_{-0.07}$\,dex), reflecting the 
heterogeneous formation histories of clusters born across the full MW 
potential.

In contrast to metallicity, [$\alpha$/Fe] is remarkably homogeneous across 
families (Fig.~\ref{Fig:kernel}, bottom), with medians spanning only 
$\sim 0.20$--$0.31$\,dex. An analysis of covariance (ANCOVA) test shows that [$\alpha$/Fe] is primarily 
driven by the [Fe/H] dependence ($p = 1.4 \times 10^{-4}$), with no 
significant residual family effect ($p \simeq 0.066$).

   \begin{figure}
        \centering
        \includegraphics[width=\hsize]{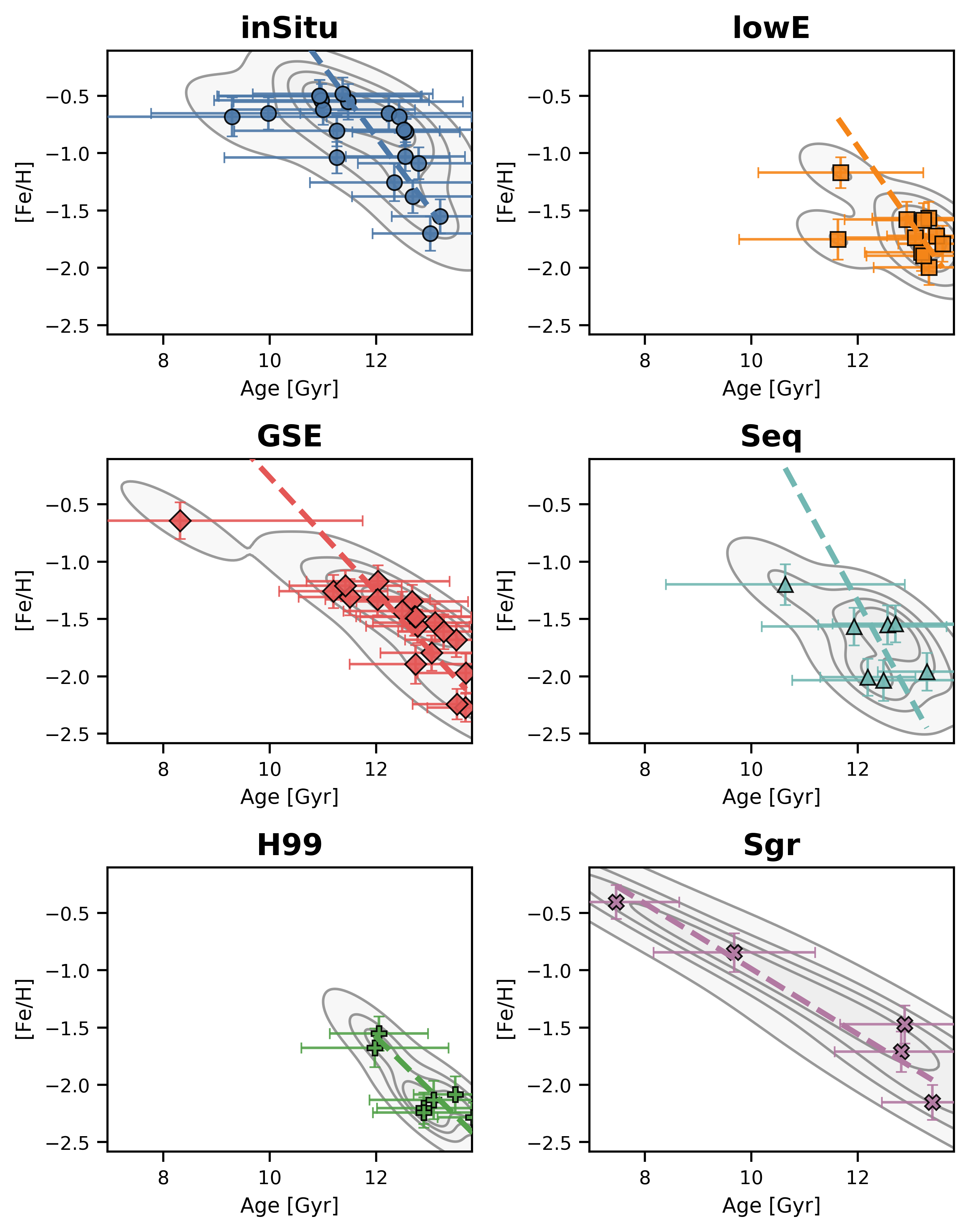}
        \caption{AMRs for the final merged progenitor groups. Each panel shows individual GCs (points) with asymmetric age and metallicity uncertainties. Grey-shaded kernel-density contours highlight the main population distribution, and the orthogonal distance regression linear fit (dashed line) accounts for errors in both variables. The linear fits provide a visual diagnostic of the overall AMR trend within each progenitor; however, they should not be interpreted as physical enrichment rates, as the inferred slopes depend sensitively on the sampled age range.}
        \label{Fig:AMR}%
    \end{figure}

\subsection{Age--metallicity relations}
\label{sec:amr_linear}

The AMRs of the final progenitor groups are shown in Fig.~\ref{Fig:AMR}, with 
orthogonal distance regression linear fits overlaid to account for 
uncertainties in both age and metallicity. Before discussing individual systems, 
we stress that these slopes depend sensitively on the sampled age range and 
cluster numbers, and should be treated as descriptive diagnostics rather than 
physical enrichment rates; the hierarchical modelling in Sect. \ref{sec:amr} 
provides the more physically motivated constraints.

With that caveat, several families show clear negative slopes indicating 
progressive enrichment. GSE has the best-defined relation 
($-0.50 \pm 0.08$\,dex\,Gyr$^{-1}$, $6.2\sigma$), spanning a broad baseline 
in both age ($\sim8.3$--$13.7$\,Gyr) and metallicity 
($[\rm Fe/H] \approx -2.27$ to $-0.64$), consistent with an extended 
pre-accretion enrichment history. H99 shows a comparable slope 
($-0.46 \pm 0.13$\,dex\,Gyr$^{-1}$, $3.5\sigma$) but over a narrower 
baseline ($\sim12.0$--$13.8$\,Gyr; $[\rm Fe/H] \approx -2.28$ to $-1.55$). 
Sagittarius is shallower ($-0.29 \pm 0.05$\,dex\,Gyr$^{-1}$, $5.7\sigma$) 
yet covers the widest metallicity range ($[\rm Fe/H] \approx -2.15$ to $-0.40$) 
and a long age baseline ($\sim7.5$--$13.4$\,Gyr), reflecting prolonged 
enrichment in a massive progenitor. The in situ population shows a 
significant slope ($-0.62 \pm 0.14$\,dex\,Gyr$^{-1}$, $4.5\sigma$) but the 
largest intrinsic dispersion ($\sigma_{\rm int} \approx 0.47$\,dex), likely 
reflecting the heterogeneous formation histories of clusters born within 
the MW potential. The lowE group yields a marginal detection 
($-0.66 \pm 0.33$\,dex\,Gyr$^{-1}$, $2.0\sigma$), while Sequoia's slope 
is unconstrained ($-0.85 \pm 0.74$\,dex\,Gyr$^{-1}$, $1.2\sigma$).

The intrinsic metallicity dispersions are equally informative. H99 is 
remarkably chemically coherent ($\sigma_{\rm int} \approx 0.08$\,dex), 
and Sagittarius, despite its wide metallicity span, shows only modest scatter 
($\sigma_{\rm int} \approx 0.13$\,dex). Both are consistent with regulated 
enrichment prior to or during disruption. GSE and lowE are intermediate 
($\sigma_{\rm int} \approx 0.3$\,dex), while the in situ population 
and Sequoia show the largest dispersions ($\sigma_{\rm int} \gtrsim 0.45$\,dex), 
reflecting heterogeneous origins in the former case and small-number statistics 
in the latter.

\begin{table*}
\caption{Posterior constraints on the AMR parameters for each accreted progenitor family.}
\label{tab:amr_posterior_summary}
\centering
\setlength{\tabcolsep}{4.5pt}
\renewcommand{\arraystretch}{1.15}
\begin{tabular}{l c c c c c c c}
\hline\hline
Group & $N_{\rm GC}$ &
$L_0$ [Gyr] & $L_{\rm stop}$ [Gyr] & $\tau$ [Gyr] &
$[\mathrm{Fe/H}]_0$ & $\Delta[\mathrm{Fe/H}]$ & $[\mathrm{Fe/H}]_{\rm stop}$ \\
\hline
lowE  & 11 &
$12.66^{+0.86}_{-0.94}$ & $7.43^{+2.24}_{-3.84}$ & $2.12^{+1.20}_{-0.76}$ &
$-1.85^{+0.10}_{-0.18}$ & $1.09^{+0.41}_{-0.34}$ & $-0.80^{+0.43}_{-0.36}$ \\
GSE   & 19 &
$13.49^{+0.19}_{-0.28}$ & $7.46^{+2.03}_{-3.88}$ & $2.06^{+0.93}_{-0.68}$ &
$-2.09^{+0.12}_{-0.10}$ & $1.32^{+0.30}_{-0.26}$ & $-0.79^{+0.28}_{-0.23}$ \\
Seq   & 7  &
$12.71^{+0.64}_{-0.76}$ & $7.48^{+2.29}_{-3.79}$ & $2.12^{+1.14}_{-0.76}$ &
$-2.02^{+0.14}_{-0.13}$ & $1.08^{+0.38}_{-0.31}$ & $-0.95^{+0.40}_{-0.29}$ \\
H99   & 7  &
$12.62^{+0.55}_{-0.54}$ & $7.51^{+2.25}_{-3.53}$ & $2.05^{+1.08}_{-0.75}$ &
$-2.17^{+0.09}_{-0.09}$ & $1.16^{+0.40}_{-0.36}$ & $-1.02^{+0.41}_{-0.36}$ \\
Sgr   & 5  &
$12.76^{+0.57}_{-0.69}$ & $6.93^{+2.01}_{-3.82}$ & $1.97^{+0.90}_{-0.69}$ &
$-1.98^{+0.16}_{-0.15}$ & $1.56^{+0.33}_{-0.28}$ & $-0.43^{+0.29}_{-0.21}$ \\
\hline
\end{tabular}
\tablefoot{We report the posterior median and the 16th--84th percentile interval.
$L_0$ and $L_{\rm stop}$ are in Gyr (look-back time), $\tau$ is in Gyr, and metallicities are in dex. Here $[\mathrm{Fe/H}]_{\rm stop} \equiv [\mathrm{Fe/H}]_0 + \Delta[\mathrm{Fe/H}]$.}
\end{table*}

  \begin{figure*}
        \centering
        \includegraphics[width=1.0\hsize]{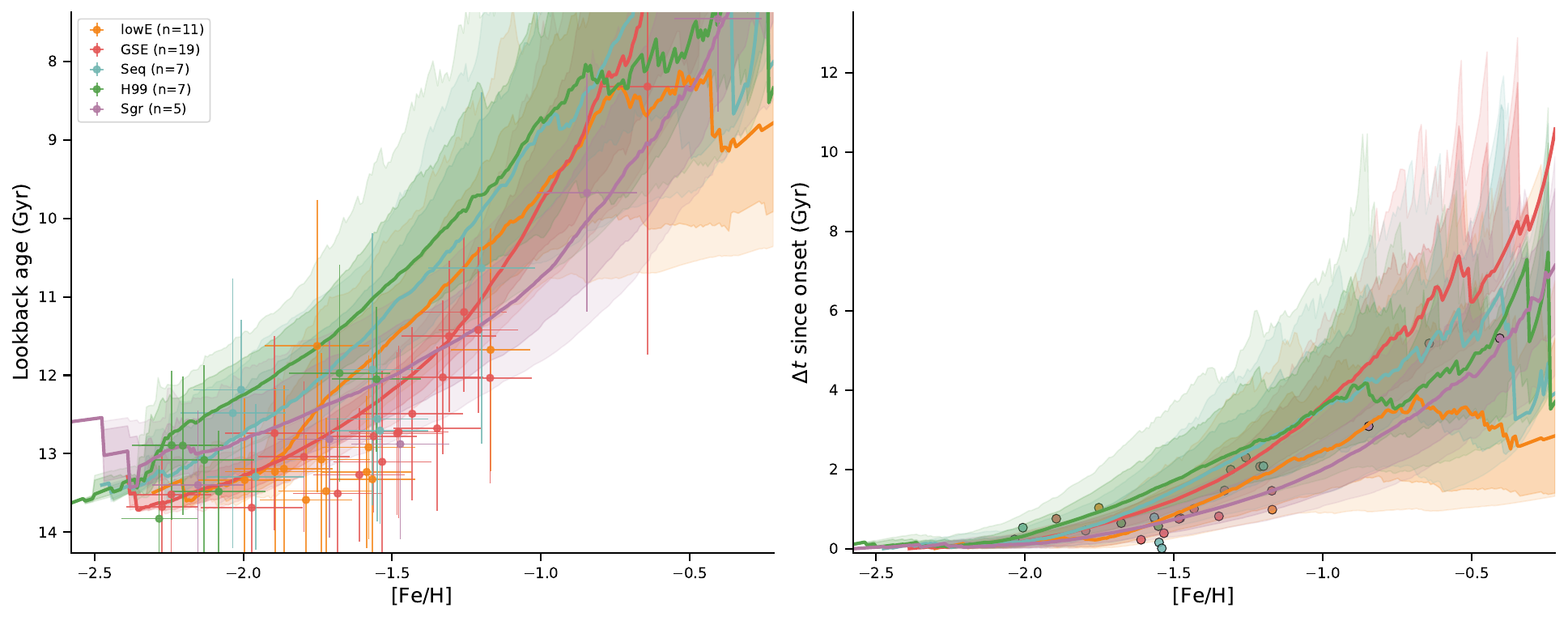}
           \caption{Hierarchical Bayesian AMRs for accreted MW progenitors traced by GCs. {\em Left:} Absolute look-back age versus $[\mathrm{Fe/H}]$, with observational uncertainties shown as error bars. Solid lines indicate the posterior median AMR for each progenitor, and shaded regions represent the 68\% (dark) and 90\% (light) credible intervals derived from posterior samples.  {\em Right:} Same relations expressed in relative time since enrichment onset, $\Delta t = L_0 - t$, highlighting the inferred enrichment duration prior to truncation.}
        \label{Fig:HIER}%
    \end{figure*}

\subsection{Hierarchical AMR constraints}
\label{sec:amr}

Posterior constraints from the hierarchical AMR model (Sect. \ref{sec:hierarc_modelling}) 
are shown in Fig.~\ref{Fig:HIER} and summarised in Table~\ref{tab:amr_posterior_summary}.

\paragraph{Enrichment timescales.}
All progenitors yield enrichment timescales in the range $\tau \sim$ 1--2\,Gyr, with the hierarchical population-level estimate $\mu_{\ln\tau} = 0.69^{+0.36}_{-0.38}$ (corresponding to $\tau \simeq$2\,Gyr). We caution, however, that this estimate carries a non-negligible prior contribution: when progenitor AMRs are fitted independently without pooling, group-level medians shift to $\tau \sim$ 1.1--1.4\,Gyr with substantially wider credible intervals, and prior sensitivity tests show that $\sim$40\% of a shift in the hyper prior mean is transmitted to the posterior. The inter-group scatter $\sigma_{\ln\tau}$ is consistent with zero,
confirming that the data cannot resolve differences in enrichment pace between systems at the present sample size. We therefore treated $\tau \lesssim$2\,Gyr as an order-of-magnitude constraint 
— while emphasising that the primary physically robust distinction between progenitors lies in the amplitude of chemical evolution $\Delta[\mathrm{Fe/H}],$ which is genuinely data-driven and insensitive to the $\tau$ prior choice\footnote{As an additional check, we fixed $\tau$ to a grid of values spanning 0.5--4\,Gyr and verified that the posteriors of $\Delta[\mathrm{Fe/H}]$ and $[\mathrm{Fe/H}]_{\rm stop}$ vary by less than 0.1\,dex, while $L_{\rm stop}$ changes by less than 0.3\,Gyr. This confirms that the inferred quantities are insensitive to the precise value of $\tau$ within its credible range.}.Initial metallicities are uniformly metal-poor across progenitors  
($\mu_{[\rm Fe/H]_0} = -2.03 \pm 0.12$, group scatter $\approx 0.21$\,dex), 
and the inferred intrinsic scatter around the fitted relations is modest 
($\sigma_{\rm int,FeH} \approx 0.08 \pm 0.05$\,dex), indicating that the 
truncated exponential captures the main AMR features without requiring 
substantial additional dispersion.

\paragraph{Enrichment amplitude and truncation.}
The main distinction between progenitors lies in the extent rather than the 
pace of chemical evolution. Most systems reach total metallicity growths of 
$\Delta[\rm Fe/H] \sim 1.1$--$1.3$\,dex before truncation, with enrichment 
rates ($R_{\rm enrich} = \Delta {\rm [Fe/H]}/ \tau$) clustering around $R_{\rm enrich} \approx 0.50$--$0.63$\,dex\,Gyr$^{-1}$ 
(GSE: $0.63^{+0.21}_{-0.14}$; H99: $0.55^{+0.26}_{-0.18}$; Seq: 
$0.51^{+0.23}_{-0.16}$; lowE: $0.50^{+0.24}_{-0.17}$). Truncation look-back 
times cluster around $L_{\rm stop} \sim 7.4$--$7.5$\,Gyr for these four systems. 
Sagittarius is the outlier in all respects: it reaches 
$\Delta[\rm Fe/H]_{\rm med} = 1.56^{+0.33}_{-0.28}$\,dex and a terminal 
metallicity of $[\rm Fe/H]_{\rm stop} = -0.43^{+0.29}_{-0.21}$, substantially 
higher than the other progenitors ($[\rm Fe/H]_{\rm stop} \simeq -1.0$ to $-0.8$), 
with a correspondingly elevated $R_{\rm enrich} = 0.79^{+0.36}_{-0.21}$\,dex\,Gyr$^{-1}$ 
and a somewhat later truncation ($L_{\rm stop} = 6.93^{+2.01}_{-3.82}$\,Gyr). 
This is consistent with more efficient metal retention and prolonged star 
formation in a comparatively massive progenitor.

\begin{figure}
    \centering
    \includegraphics[width=\columnwidth]{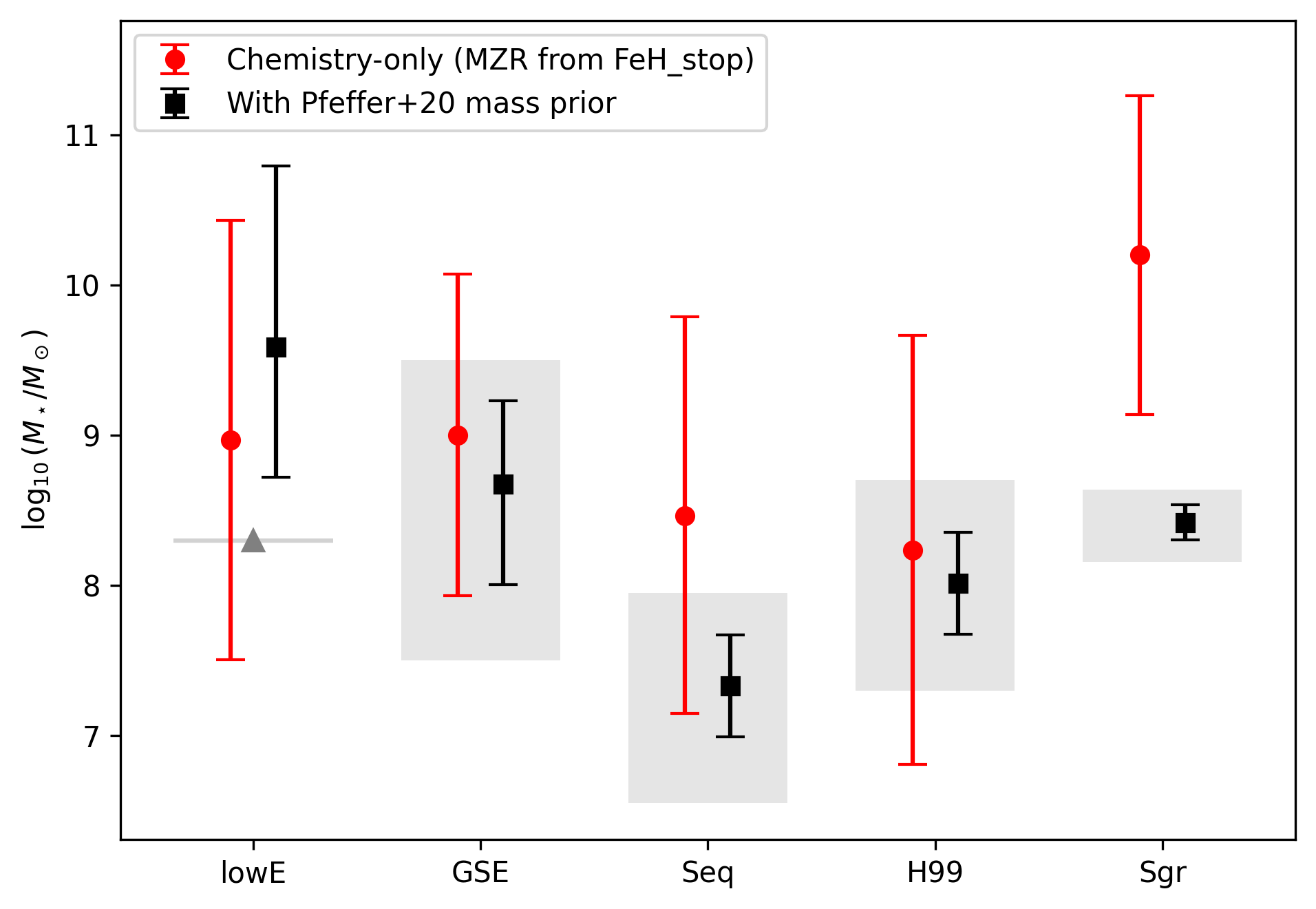}
    \caption{Progenitor stellar masses inferred from the AMR subsample compared to dynamical constraints. Grey-shaded boxes indicate the allowed progenitor stellar-mass ranges inferred from GC orbital information in {\tt E-MOSAICS} by \citet{pfeffer20}. Black points show the chemistry-only $M_\star$ posteriors obtained by inverting the dwarf mass--metallicity relation using the AMR-inferred $[\mathrm{Fe/H}]_{\rm stop}$ (error bars are the 16th--84th percentiles from MC propagation). Red points show the corresponding importance-weighted posteriors after applying the \citet{pfeffer20} mass constraints as a prior in $\log M_\star$.}
    \label{fig:mstar_pfeffer}
\end{figure}

\paragraph{Progenitor mass estimates.}
The AMR constraints can be combined with the GC census to estimate progenitor 
masses. Anchoring cluster counts to the full 154-object catalogue of 
\citet{massari19,massari23} and applying the empirical $N_{\rm GC}$--$M_{\rm vir}$ 
relation \citep{harris17,burkert20,reina22}, we infer halo masses of 
$M_{\rm vir} \sim (3$--$15) \times 10^{10}\,{\rm M_\odot}$. Translating 
AMR-inferred terminal metallicities into stellar masses via the dwarf-galaxy 
mass--metallicity relation \citep{kirby13} yields $M_\star \sim 10^9\,{\rm M_\odot}$ 
for GSE and the low-energy group, lower values for Sequoia and the Helmi 
progenitor, and a substantially higher value for Sagittarius. The chemistry-only 
estimate for Sagittarius, however, exceeds the dynamical constraint of 
$M_\star \sim (2$--$3) \times 10^8\,{\rm M_\odot}$ from \citet{pfeffer20}, 
likely because the high terminal metallicity of its GC system does not 
straightforwardly trace the global field-star mass--metallicity relation in 
a system with prolonged and complex chemical evolution. Incorporating the 
\citet{pfeffer20} dynamical constraints as priors shifts Sagittarius into the 
expected mass range while leaving the GC-number-based halo masses essentially 
unchanged (Fig.~\ref{fig:mstar_pfeffer}), and identifies GSE and the low-energy progenitor as the dominant accretion events.

\subsection{MPs and environment}
\label{sec:MP_environment}

Having established the large-scale enrichment histories of the progenitor systems, we turned to their internal stellar populations, asking whether the helium-based MP indicators that were held fixed in the AMR modelling carry any residual imprint of the galactic environment in which each cluster formed. In particular, using the regression framework described in Sect. \ref{sec:MP_methods_MPs}, 
we investigated whether MP properties depend on progenitor environment after 
controlling for cluster mass and metallicity.

Cluster mass is the dominant predictor across all three indicators. The helium 
spread $\delta Y$ increases with mass ($p\approx10^{-5}$), the first-population 
fraction $f_{\rm P1}$ decreases ($p\approx3\times10^{-5}$), and the mean helium 
abundance $\bar{Y}$ shows a weaker but significant positive dependence 
($p\approx0.02$). Metallicity does not contribute significantly once mass is 
included ($p>0.1$). These mass scalings are consistent across both in situ and 
accreted systems: interaction tests find no evidence that the slope of the 
mass--MP relation differs between progenitors, indicating that this relation is 
universal.

\begin{figure}
\centering
\includegraphics[width=\hsize]{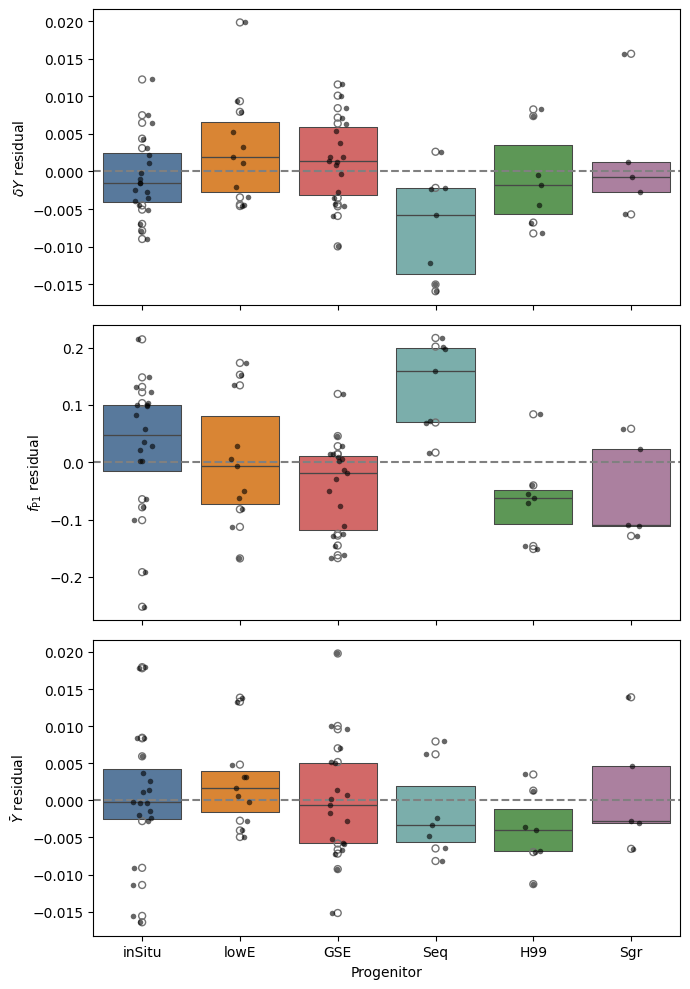}
\caption{Residuals of the multiple-population properties $\delta Y$ 
({\em top}), $f_{\mathrm{P1}}$ ({\em middle}), and $\bar{Y}$ ({\em bottom}), obtained 
after subtracting the mass and metallicity trends, grouped by progenitor family. The dashed horizontal line 
marks zero residual, i.e.\ agreement with the global 
mass--metallicity relation; negative values indicate clusters lying 
below the predicted trend and do not correspond to 
unphysical negative helium enhancements (the physical $\delta Y$ 
is non-negative by construction). Boxes span the interquartile range, and horizontal lines mark the median. Individual clusters are shown as empty 
circles; filled points are the same clusters slightly shifted for 
visualisation.}
\label{fig:MP_residuals_prog}
\end{figure}

After removing the mass and metallicity dependence, we compared the residual MP 
properties across progenitor families (Fig.~\ref{fig:MP_residuals_prog}). We emphasise that the quantities plotted in 
Fig.~\ref{fig:MP_residuals_prog} are residuals of $\delta Y$, 
$f_{\mathrm{P1}}$, and $\bar{Y}$ after the mass and metallicity 
dependences have been 
removed, rather than the physical MP properties themselves. By 
construction, the observed helium spread satisfies $\delta Y \geq 0$, 
since it measures the enhancement of the second population 
relative to the primordial one. The residual $\delta Y$ shown here, 
however, represents the deviation of a given cluster from the mean 
mass--metallicity trend and can therefore take either sign: a negative 
residual simply indicates a cluster whose helium spread is 
smaller than predicted by the global 
$\delta Y(M_{\star},\mathrm{[Fe/H]})$ relation, not a physically 
negative helium enhancement. The same interpretation applies to the 
$f_{\mathrm{P1}}$ and $\bar{Y}$ panels.

The mean helium abundance ($\bar{Y}$) shows no significant progenitor dependence. The 
helium spread $\delta Y$ shows a nominally significant progenitor term 
($p \approx 0.03$), driven by slightly smaller spreads in Sequoia clusters, but 
this signal does not survive removal of the two most uncertain Sequoia members 
(Palomar~15 and Pyxis, which carry ambiguous or conflicting literature 
classifications; see Table~\ref{tab:cluster_summary}): dropping either cluster 
individually pushes $p > 0.08$, and removing both yields $p = 0.26$. We 
therefore treat the $\delta Y$--Sequoia trend as suggestive but unconfirmed.

\begin{figure}
\centering
\includegraphics[width=\hsize]{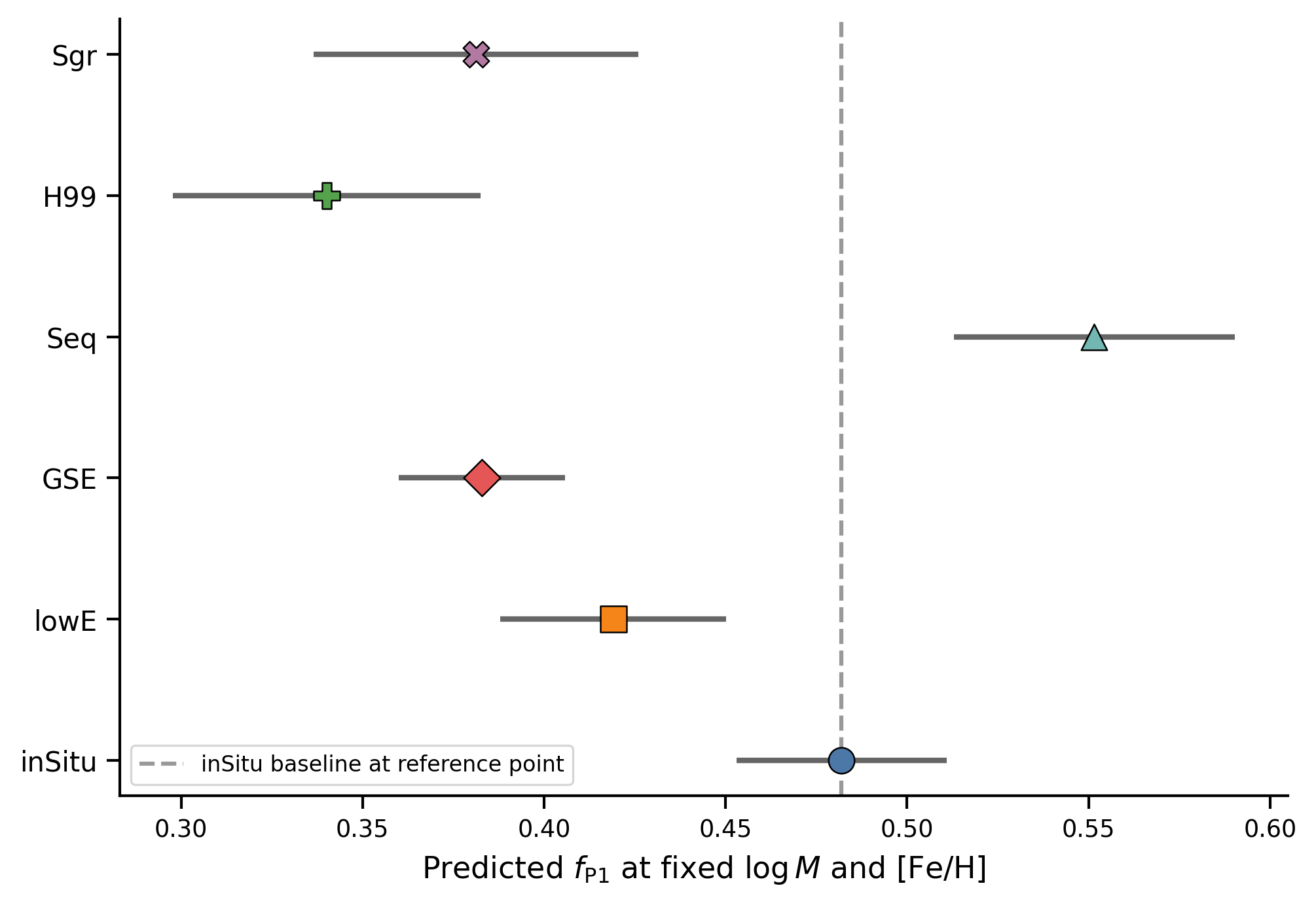}
\caption{Predicted values of the linear model evaluated at a common reference point ($\log M_z = 0$, $[\mathrm{Fe/H}]_z = 0$) for each progenitor family. Points represent the estimated marginal means, and horizontal bars indicate the $68\%$ confidence intervals derived from the fitted model. After controlling for cluster mass and metallicity, the fraction of first-population stars ($f_{\rm P1}$) shows systematic differences between progenitors, with Sequoia clusters exhibiting higher $f_{\rm P1}$ values at fixed mass and metallicity.}
\label{fig:MP_offset}
\end{figure}

In contrast, the $f_{\rm P1}$--Sequoia signal is robust. Sequoia clusters show 
systematically higher $f_{\rm P1}$ at fixed mass and metallicity 
($p \approx 8.6\times10^{-4}$; Fig.~\ref{fig:MP_offset}), corresponding to smaller enriched stellar fractions. This result strengthens when 
Palomar~15 is removed ($p \approx 3.9\times10^{-4}$) and remains significant 
across all reclassification subsets, including after dropping both borderline 
members ($p \approx 1.4\times10^{-3}$, $N_{\rm Seq}=5$). The direction and 
magnitude of the Sequoia coefficient are stable throughout, indicating that the 
signal is a property of the group rather than an artefact of uncertain 
classification. Among the accreted families, GSE and H99 clusters tend towards 
lower $f_{\rm P1}$ at fixed mass and metallicity, suggesting that the relative 
proportions of enriched and non-enriched stars retain a weak but real imprint 
of the formation environment.

As an independent check, we modelled chromosome-map pseudo-colour widths 
\citep{milone17} for the 56 clusters with available measurements. Both 
$W_{C~F275W,F336W,F438W}$ and $W_{mF275W-mF814W}$ are strongly predicted by 
cluster mass and metallicity ($p \lesssim 10^{-7}$), while progenitor origin 
and cluster age contribute no additional explanatory power once these are 
included. This confirms that the mass--MP scaling is universal and 
environmentally blind, consistent with the helium-based results above.

\subsection{Dynamical evolution tests}
\label{sec:MP_dynamics}

Before interpreting the progenitor dependence of $f_{\rm P1}$ as a formation signal, we assessed whether any of the detected MP trends could instead reflect long-term dynamical erosion of MPs over the clusters' orbital histories.
In particular, we tested whether MP tracers are 
modified by long-term dynamical evolution, examining correlations with 
present-day orbital parameters, mass-loss proxies, and eccentricity-binned 
population fractions.

Spearman rank tests against orbital quantities reveal only weak trends. The 
strongest involve pericentric distance, where $\delta Y$ and $\bar{Y}$ 
show mild anti-correlations ($p=0.024$ and $p=0.029$), but neither survives 
false-discovery-rate correction ($p_{\rm FDR}\approx0.06$). All other orbital 
quantities yield $|\rho|\lesssim0.21$ with no significant residuals after 
false-discovery-rate correction.

Mass-loss proxies derived from initial mass estimates \citep{baumgardt18} —
the logarithmic parameter $\Delta\log M = \log_{10}(M_{\rm init}/M_{\rm now})$
and fractional mass loss $f_{\rm loss}$ — show apparently significant
correlations with all three MP indicators (raw Spearman $p \leq 0.015$).
However, because current and initial cluster masses are strongly correlated,
these signals disappear entirely once present-day cluster mass and metallicity
are included as covariates: the mass-loss term is non-significant for all
indicators ($p = 0.22$, $0.35$, and $0.91$ for $\delta Y$, $f_{\rm P1}$,
and $\bar{Y}$, respectively), confirming that the apparent signal is a
secondary reflection of the mass--MP relation rather than evidence of
dynamical erosion.

Two further null results reinforce this conclusion. Dividing clusters within 
each progenitor into high- and low-eccentricity bins yields no significant 
difference in median $f_{\rm P1}$ for any group with sufficient members 
(in situ: $p=0.40$; GSE: $p=0.42$; lowE: $p=0.63$), arguing against 
preferential erosion of first-population stars on radial orbits. Similarly, 
present-day cluster mass distributions do not differ significantly between 
progenitor groups 
(all pairwise 
Kolmogorov–Smirnov tests yield $p_{\rm FDR}\gtrsim0.33$), providing no evidence that tidal 
destruction has systematically biased the surviving cluster population of 
any particular progenitor.

\section{Discussion and conclusions}\label{sec:discussion}

\subsection{Chemical evolution of the MW progenitors}

The metallicity distributions of the six dynamical families broadly follow 
the galaxy stellar mass--metallicity relation \citep{kirby13,kruijssen19,massari19,forbes20}: lower-mass progenitors produce more metal-poor GC 
systems, while the in situ population is the most enriched. Helmi-like 
clusters occupy the most metal-poor regime; GSE, Sequoia, and lowE sit at 
intermediate metallicities; and Sagittarius spans the widest range with the 
largest intrinsic dispersion~\citep{bellazzini06,siegel07}, consistent with more efficient metal retention and sustained star formation in a comparatively massive progenitor \citep{hasselquist17, ruizlara20}.

Despite these [Fe/H] differences, $[\alpha/\rm Fe]$ ratios are largely 
homogeneous across families once the metallicity dependence is removed, with 
no significant residual offsets between progenitors. This indicates that virtually all clusters in the sample formed during the early $\alpha$-enhanced phase of chemical evolution, before substantial Type~Ia supernova enrichment, and that the star-formation histories of these progenitors were broadly comparable in their early phases despite their different masses and accretion epochs \citep{venn04,matteucci12}

\subsection{Age--metallicity relations and enrichment histories}

 The hierarchical AMR modelling places enrichment timescales broadly in the range $\tau \lesssim$2\,Gyr across all progenitors. This convergence is physically plausible — it is consistent with gas-consumption timescales in early star-forming dwarf galaxies \citep{matteucci01,lanfranchi03} — but we caution that it is not purely data-driven. We therefore treat $\tau \lesssim$2\,Gyr as an order-of-magnitude upper constraint and caution against over-interpreting the precise value.

The main distinction between systems lies in the extent of their chemical evolution: most reach $\Delta[\rm Fe/H] \sim 1.1$--$1.3$\,dex before truncation, while Sagittarius achieves $\Delta[\rm Fe/H] \simeq 
1.6$\,dex and substantially higher terminal metallicities, consistent with 
more efficient metal retention and sustained star formation in a more massive 
progenitor \citep{hasselquist17}. These differences are consistent with the galaxy stellar mass--metallicity relation \citep{kirby13}, and imply that the progenitors contributing to the MW
halo spanned stellar masses of roughly $10^{7}--10^{9}\,M_\odot$, consistent
with expectations from cosmological simulations of galaxy assembly.

Truncation epochs cluster around look-back 
times of $\sim 7$--$8$\,Gyr for most systems, in broad agreement with the 
accretion epochs inferred from dynamical studies and cosmological simulations 
\citep{belokurov18,helmi18,pfeffer20,massari26}. Satellite galaxies lose their gas and cease star formation shortly after accretion into a larger halo through tidal stripping and ram-pressure
processes, so this coincidence between chemical truncation and dynamically
inferred accretion epochs provides an independent confirmation of the
hierarchical assembly scenario.

Terminal metallicities inferred from the AMR provide an approximate progenitor mass
ranking, though the ordering depends on whether purely chemical or dynamical
scalings are applied. Chemistry alone overestimates the Sagittarius mass
because its GC system reaches anomalously high metallicities relative to its
true stellar mass — a tension that is physically informative. The high
terminal metallicity of the Sagittarius GC system likely reflects a
combination of localised enrichment in the dense central regions of the
progenitor and the fact that GCs preferentially sample the metal-rich tail
of the field-star metallicity distribution in a system with complex,
multi-episode star formation \citep{ruizlara20, hasselquist17}. This makes
the dwarf-galaxy mass--metallicity relation \citep{kirby13} a poor mass
estimator for Sagittarius specifically, and serves as a caution against
applying chemical scalings uniformly across progenitors with very different
star-formation histories. Incorporating the {\tt E-MOSAICS} dynamical
constraints \citep{pfeffer18, pfeffer20} resolves this tension and identifies
GSE and the low-energy progenitor as the dominant events~\citep{aguado25, massari26}. This is consistent with
a $\Lambda$CDM assembly history featuring two major mergers --- a deep early
event at $z \gtrsim 2$ and a later GSE-like accretion at $z \sim 1$--$2$ ---
superimposed on numerous minor accretions \citep{kruijssen19}.

\subsection{Universality of cluster-scale physics}

In contrast to the clear environmental signatures in GC ages and metallicities,
MP properties are largely insensitive to progenitor origin. After controlling
for cluster mass and metallicity, neither $\delta Y$ nor $\bar{Y}$ shows a
significant dependence on dynamical family, and the mass--MP scaling relation
is statistically indistinguishable across in situ and accreted systems. These
findings are confirmed by independent chromosome-map pseudo-colour widths,
which show no progenitor dependence once mass and metallicity are included
\citep{milone17,milone20,bastian18}.

The mass scalings of $\delta Y$ and $f_{\rm P1}$ are qualitatively consistent
with the inertial-inflow model of \citet{gieles25}, in which a supermassive
star forming via stellar collisions at the cluster centre pollutes the
intracluster medium through its winds, with the mass fraction of chemically
processed stars scaling with cluster potential depth. That model also predicts
a dependence of the anomalous-star fraction on metallicity; our data do not
independently confirm this, likely because mass
and metallicity are correlated in our sample and mass absorbs the combined
signal. The high-redshift environment in which most GCs formed ---
characterised by denser, more pressurised interstellar media and higher star
formation efficiencies \citep{Trenti, kruijssen12,kruijssen19,pfeffer20} --- provides
a natural setting for this behaviour, with differences between host galaxies
manifesting primarily in their chemical enrichment histories and merger
timescales rather than in the internal abundance patterns of individual
clusters.

Sequoia clusters show systematically higher
first-population fractions at fixed mass and metallicity, and this signal is
robust to the removal of borderline members (Sect. \ref{sec:MP_environment}).
The robustness of this result across both our probabilistic classification and
the independent assignments of \citet{massari19, massari23} confirms that it
reflects a genuine physical property of the Sequoia progenitor rather than an
artefact of any particular classification methodology. This residual
environmental dependence has no counterpart in the \citet{gieles25} framework,
which operates purely at cluster scales, and may reflect a secondary modulation
of the enrichment balance by large-scale conditions in the Sequoia progenitor
that their model does not yet capture. Among the accreted families, GSE and H99
clusters tend towards lower $f_{\rm P1}$ at fixed mass and metallicity, further
supporting the interpretation that the relative proportions of enriched and
non-enriched stars retain a weak but real imprint of the formation environment,
even while the overall amplitude of helium enrichment does not.

We interpret the dominant mass dependence as evidence that the physical
conditions required to produce light-element abundance variations were
widespread across galaxies in the early Universe and regulated primarily by the
cluster gravitational potential, operating in a largely universal manner
regardless of host galaxy. The sole robust exception --- a residual progenitor
dependence in $f_{\rm P1}$ --- suggests that at least one aspect of the
enrichment balance is sensitive to the large-scale environment \citep{martocchia18,milone20}, pointing towards
a secondary role for host-galaxy conditions that future models will need to
accommodate.

\subsection{Dynamical evolution and robustness of MP signatures}

The environmental blindness of helium enrichment amplitude would be undermined if tidal evolution systematically modified MP signatures over a Hubble time; we therefore verified that the mass-scaling results are not a secondary consequence of preferential mass loss in particular progenitor systems. Correlations between MP tracers and orbital parameters are weak and
statistically insignificant after correction for multiple testing. Mass-loss
proxies show apparent correlations with MP indicators, but these vanish once
cluster mass is included in the models, confirming that the signal is a
secondary reflection of the mass--MP relation rather than evidence of
dynamical erosion \citep{baumgardt18}. Helium-related MP signatures therefore
appear to be set at formation and largely preserved over $\sim$10\,Gyr of
dynamical evolution, consistent with the observational conclusion of
\citet{bastian15} that tidal mass loss does not drive the present-day
distribution of enriched-star fractions across the Galactic GC population.

\subsection{Implications for GC formation and Galactic assembly}

The Galactic GC system encodes two largely distinct layers of information.
Ages, metallicities, and orbital properties trace the hierarchical assembly
and chemical enrichment of their progenitor galaxies: different systems
experienced similar enrichment timescales but different extents of chemical evolution, and
the AMRs provide indirect constraints on progenitor masses and merger epochs
that are broadly consistent with the $\Lambda$CDM picture of MW assembly.
Satellite galaxies lose their gas and cease star formation shortly after
accretion into a larger halo due to tidal stripping and ram-pressure processes,
so the coincidence between the chemical truncation epochs inferred from the
AMRs and the dynamically inferred accretion times provides an independent
confirmation of the hierarchical assembly scenario. In contrast, the internal
complexity of MPs is primarily governed by cluster mass and is largely
insensitive to galactic environment, with the sole robust exception being the
residual progenitor dependence in $f_{\rm P1}$.
This duality --- environmental memory in the ages and metallicities,
near-universal cluster-scale physics in the stellar populations --- highlights
the complementary power of chemo-dynamical classification and homogeneous MP
modelling as probes of both galaxy assembly and star cluster formation.

A key methodological novelty of this work is the explicit treatment of MPs in the stellar parameter inference \citep{valcin26}: by modelling two coexisting populations in each CMD, we avoided the systematic helium-driven age biases that affect single-population analyses \citep{salaris05} and obtained helium-specific MP indicators --- $\delta Y$, $\bar{Y}$, and $f_{\rm P1}$ --- that can be directly compared to theoretical predictions. This combination of MP-aware ages and multi-progenitor classification, applied to a homogeneous 69-cluster sample, represents the first test of whether helium enrichment properties carry an environmental imprint across the full hierarchy of MW accretion events. As future surveys extend precise age and abundance measurements to larger samples
of clusters in nearby galaxies, this combination of approaches may become an
increasingly powerful tool for reconstructing merger histories beyond the
MW.

\begin{acknowledgements}
This work has made use of data from the European Space Agency (ESA)
mission \textit{Gaia}\footnote{https://www.cosmos.esa.int/gaia},
processed by the \textit{Gaia} Data Processing and Analysis Consortium
(DPAC). Funding for the DPAC has been provided by national institutions,
in particular the institutions participating in the \textit{Gaia}
Multilateral Agreement.

This research has made use of the SIMBAD database, operated at CDS,
Strasbourg, France, and NASA’s Astrophysics Data System.

All computations were performed in Python~3.12 using the scientific
Python ecosystem, including \texttt{NumPy},
\texttt{SciPy},
\texttt{pandas},
\texttt{scikit-learn},
\texttt{statsmodels},
\texttt{matplotlib},
and \texttt{seaborn}. Orbit integrations and dynamical quantities
were computed with \texttt{galpy} \citep{bovy15}. Statistical analyses
and probabilistic modelling were implemented using standard scientific
Python tools.
\end{acknowledgements}

\bibliographystyle{aa}
\bibliography{references.bib}

\begin{appendix}\label{app}

\section{Assessment of probabilistic cluster classifications}
\label{sec:appendix_classification}

\nolinenumbers

Table~\ref{tab:cluster_summary} presents the full probabilistic classification. 
Each cluster is assigned to the family with the highest posterior probability 
$p_{\rm max}$, graded into secure ($p_{\rm max} \geq 0.80$, 
$\Delta p \geq 0.20$), \textit{likely} ($p_{\rm max} \geq 0.60$, 
$\Delta p \geq 0.10$), and ambiguous tiers based on the probability 
margin $\Delta p = p_{\rm max} - p_{\rm 2nd}$. These conservative thresholds 
ensure that secure assignments require both a high posterior probability and a 
clear separation from neighbouring families.

Most clusters satisfy the secure criterion. Sagittarius, Sequoia, H99, and 
in situ clusters are almost entirely secure, reflecting the compact 
or dynamically distinct loci of these systems. GSE shows the largest fraction 
of likely or ambiguous assignments, consistent with the expected overlap of 
early accretion debris in the inner halo. Ambiguous objects — a small minority 
— occupy transitional phase-space regions where debris from multiple events 
has mixed. The classification is stable under moderate variations of the 
feature set and covariance regularisation; differences under perturbation 
occur almost exclusively among objects already flagged as ambiguous.

We note that the multivariate Gaussian assumption is a deliberate simplification: real tidal debris streams have non-Gaussian distributions in action space, particularly for Sagittarius and Sequoia. However, the strong covariance regularisation and soft priors applied here mean that the classifier effectively acts as a smoothed nearest-neighbour assignment rather than a strict parametric model, and the validation tests above confirm that non-Gaussian features do not substantially degrade classification accuracy for the families relevant to the MP analysis.

\subsection{Quantitative validation}
\label{sec:validation}

We assessed the robustness and physical plausibility of the classification 
through three complementary tests.

\paragraph{Phase-space coherence.}
For each family we computed the hypervolume, $V = \sqrt{\det(\Sigma)}$, in 
standardised $(E_{J}, L_z, L_\perp, J_r, [\rm Fe/H])$ space 
(Table~\ref{tab:phase_space_volume}). The resulting hierarchy — 
Sgr $\ll$ lowE/in situ $\ll$ H99 $\lesssim$ GSE $\lesssim$ Seq — reflects 
physical expectations: Sagittarius is by far the most compact 
($\log_{10}V = -10.25$), more than eight orders of magnitude smaller than the 
other families, consistent with its origin in a recent, not-yet-phase-mixed 
accretion event. At the other extreme, Sequoia and GSE show the largest 
volumes ($\log_{10}V = -1.91$ and $-2.14$), consistent with early, massive 
mergers whose debris has dispersed broadly in action space. The in 
situ and lowE families occupy intermediate volumes and similar binding 
energies, but their dynamical proximity likely reflects the restricted phase 
space accessible to inner-halo debris rather than a common origin: recent 
work interprets lowE as the remnant of an early massive merger now strongly 
phase-mixed deep in the MW potential \citep{massari26}. The large condition 
number of the Sagittarius covariance matrix further reflects the near-linear, 
stream-like alignment of its debris; the moderate values for other families 
indicate well-resolved, more isotropic distributions.

\paragraph{Comparison with literature classifications.}
Comparison against CARMA \citep{massari19,massari23}\footnote{Available at \url{https://www.oas.inaf.it/it/ricerca/m2-it/carma-it/}} via a row-normalised 
confusion matrix shows strong diagonal dominance: Sagittarius and lowE are 
recovered with perfect consistency; H99 at 86\%; GSE at 84\% (with limited 
leakage into H99 and Seq due to known inner-halo overlap); and Sequoia at 
83\%. Global metrics yield an Adjusted Rand Index (ARI) of 0.595 and a Normalized Mutual Information (NMI) of 0.728; a  probability-weighted agreement score of 0.473 confirms that discrepancies 
are confined to low-confidence transition objects. Agreement with 
\citet{youakim25} is equally strong: GSE-related subgroups (G2, sg-3/4/5) 
are recovered at $p_{\rm max} \gtrsim 0.9$; the Sequoia subgroup (G4, sg-8) 
is cleanly identified; Sagittarius is perfectly separated; and G1 
`Splash' clusters map preferentially to our lowE family, consistent with 
an early heated-population interpretation.

\paragraph{Inter-family separability.}
Mahalanobis distances between family centroids confirm the picture above. 
Sagittarius is strongly isolated ($d > 90$ from GSE and H99). GSE, H99, and 
lowE partially overlap ($d \approx 4$--5), and the in situ--lowE 
separation is the smallest in the sample ($d \sim 2.7$), reflecting their 
shared inner-halo phase space rather than a common progenitor. Sequoia sits 
at intermediate separations from GSE and H99 ($d \approx 5$--8), consistent 
with its retrograde but partially overlapping location. The global silhouette 
score of 0.167 is moderate but physically expected: long-term phase mixing 
erases sharp boundaries between early accretion structures, so clean 
separation is not anticipated.

Taken together, these tests confirm that the classification is statistically 
robust and physically consistent with the known dynamical structure of the 
MW GC system.

\begin{table}
\caption{Phase-space hypervolume of each dynamical family in standardised action--energy space.}
\label{tab:phase_space_volume}
\centering
\begin{tabular}{lrrrr}
\hline\hline
Family & $N$ & $\log_{10} V$ & $V$ & Condition number \\
\hline
Sgr     & 5  & $-10.25$ & $5.67\times10^{-11}$ & $3.3\times10^{16}$ \\
lowE    & 11 & $-4.16$  & $6.86\times10^{-5}$  & $2.2\times10^{2}$ \\
inSitu  & 20 & $-3.62$  & $2.40\times10^{-4}$  & $1.6\times10^{2}$ \\
H99     & 7  & $-2.53$  & $2.93\times10^{-3}$  & $6.2\times10^{2}$ \\
GSE     & 19 & $-2.14$  & $7.20\times10^{-3}$  & $2.4\times10^{2}$ \\
Seq     & 7  & $-1.91$  & $1.24\times10^{-2}$  & $2.1\times10^{4}$ \\
\hline
\end{tabular}
\tablefoot{The volume is defined as $\sqrt{\det(\Sigma)}$, where $\Sigma$ is the covariance matrix of 
($E_{J,\mathrm{int}}, L_z, L_\perp, J_r, \mathrm{[Fe/H]}$). The extreme condition number for Sagittarius reflects the small sample size and near-linear alignment of its debris in action space; the volume estimate should therefore be interpreted qualitatively rather than as a precise geometric measurement.}
\end{table}

\longtab[1]{
\begin{longtable}{llrlll}
\caption{Probabilistic dynamical classification of Galactic GCs.}\\
\label{tab:cluster_summary}\\
\hline\hline
Cluster & Class & $p_{\max}$ & TIER & CARMA & YL25 (coarse/fine) \\
\hline
\endfirsthead
\caption{continued.}\\
\hline\hline
Cluster & Class & $p_{\max}$ & TIER & CARMA & YL25 (coarse/fine) \\
\hline
\endhead
\hline
\endfoot

Pal~1 & GSE & 1.000 & secure & GSE & G-5/sg-16 \\
NGC~6981 & GSE & 0.987 & secure & GSE & G-2/sg-4 \\
NGC~2298 & GSE & 0.977 & secure & GSE & G-2/sg-5 Pontus \\
NGC~4147 & GSE & 0.974 & secure & GSE & G-2/sg-3 \\
NGC~6779 & GSE & 0.973 & secure & GSE & G-2/sg-5 Pontus \\
NGC~7089 & GSE & 0.966 & secure & GSE & G-2/sg-4 \\
NGC~1261 & GSE & 0.963 & secure & GSE & G-2/sg-4 \\
NGC~6341 & GSE & 0.963 & secure & GSE & G-2/sg-5 Pontus \\
NGC~1851 & GSE & 0.958 & secure & GSE & G-2/sg-4 \\
NGC~5286 & GSE & 0.942 & secure & GSE & G-2/sg-4 \\
NGC~2808 & GSE & 0.940 & secure & GSE & G-2/sg-4 \\
NGC~6934 & GSE & 0.936 & secure & H-E & G-3/sg-7 Candidate Merger \\
NGC~0362 & GSE & 0.899 & secure & GSE & G-2/sg-4 \\
NGC~7099 & GSE & 0.899 & secure & GSE & G-2/sg-5 Pontus \\
NGC~0288 & GSE & 0.797 & likely & M-D & G-3/sg-6 Thamnos \\
NGC~5904 & GSE & 0.756 & likely & GSE & G-2/sg-3 \\
NGC~6584 & GSE & 0.682 & likely & H99 & G-2/sg-3 \\
NGC~5139 & GSE & 0.623 & likely & GSE & G-3/sg-6 Thamnos \\
NGC~6205 & GSE & 0.510 & ambiguous & GSE & G-8/Pre-disc \\
NGC~5053 & H99 & 1.000 & secure & H99 & G-5/sg-12 LMS-1/Wukong \\
NGC~5024 & H99 & 1.000 & secure & H99 & G-5/sg-11 \\
NGC~4590 & H99 & 1.000 & secure & H99 & G-5/sg-15 \\
Rup~106 & H99 & 0.939 & secure & H99 & G-5/Accreted Structures \\
NGC~5272 & H99 & 0.815 & secure & H99 & G-5/sg-10 Helmi Streams \\
NGC~6426 & H99 & 0.809 & secure & GSE & G-5/sg-10 Helmi Streams \\
NGC~7078 & H99 & 0.666 & likely & H99 & G-5/sg-10 Helmi Streams \\
NGC~6101 & Seq & 1.000 & secure & Seq & G-4/sg-8 Sequoia \\
NGC~3201 & Seq & 1.000 & secure & Seq & G-4/sg-8 Sequoia \\
NGC~5466 & Seq & 1.000 & secure & Seq & G-3/sg-7 Candidate Merger \\
Ic~4499 & Seq & 0.998 & secure & Seq & G-4/sg-8 Sequoia \\
Pyxis & Seq & 0.992 & secure & Elq & G-5/sg-14 Sagittarius \\
NGC~7006 & Seq & 0.978 & secure & Seq & G-2/sg-1 \\
Pal~15 & Seq & 0.571 & ambiguous & GSE & G-3/sg-7 Candidate Merger \\
Ter~7 & Sgr & 1.000 & secure & Sgr & G-5/sg-14 Sagittarius \\
Pal~12 & Sgr & 1.000 & secure & Sgr & G-5/sg-14 Sagittarius \\
Ter~8 & Sgr & 1.000 & secure & Sgr & G-5/sg-14 Sagittarius \\
Arp~2 & Sgr & 1.000 & secure & Sgr & G-5/sg-14 Sagittarius \\
NGC~6715 & Sgr & 1.000 & secure & Sgr & G-5/sg-14 Sagittarius \\
NGC~6624 & inSitu & 1.000 & secure & M-B & G-6/Bulge \\
NGC~6304 & inSitu & 1.000 & secure & M-B & G-7/Post-disc \\
NGC~6637 & inSitu & 1.000 & secure & M-B & G-7/Post-disc \\
NGC~6441 & inSitu & 1.000 & secure & M-D & G-8/Pre-disc \\
NGC~6352 & inSitu & 1.000 & secure & M-D & G-7/Post-disc \\
Lynga~7 & inSitu & 1.000 & secure & M-D & G-7/Post-disc \\
NGC~6388 & inSitu & 1.000 & secure & M-B & G-8/Pre-disc \\
NGC~6496 & inSitu & 1.000 & secure & M-D & G-8/Pre-disc \\
NGC~6652 & inSitu & 0.999 & secure & M-B & G-8/Pre-disc \\
NGC~6366 & inSitu & 0.999 & secure & M-D & G-8/Pre-disc \\
NGC~5927 & inSitu & 0.999 & secure & M-D & G-7/Post-disc \\
NGC~6362 & inSitu & 0.990 & secure & M-D & G-7/Post-disc \\
NGC~6838 & inSitu & 0.983 & secure & M-D & G-7/Post-disc \\
NGC~6171 & inSitu & 0.974 & secure & M-B & G-7/Post-disc \\
NGC~0104 & inSitu & 0.970 & secure & M-D & G-7/Post-disc \\
NGC~6723 & inSitu & 0.969 & secure & M-B & G-7/Post-disc \\
NGC~6717 & inSitu & 0.871 & secure & M-B & G-7/Post-disc \\
NGC~6752 & inSitu & 0.763 & likely & M-D & G-7/Post-disc \\
NGC~6656 & inSitu & 0.679 & likely & M-D & G-8/Pre-disc \\
NGC~6218 & inSitu & 0.563 & ambiguous & M-D & G-1/Splash \\
NGC~6093 & lowE & 0.961 & secure & low-E & G-1/Splash \\
NGC~6144 & lowE & 0.953 & secure & low-E & G-1/Splash \\
NGC~6541 & lowE & 0.947 & secure & low-E & G-1/Splash \\
NGC~5986 & lowE & 0.939 & secure & low-E & G-8/Pre-disc \\
NGC~6535 & lowE & 0.936 & secure & Seq & G-1/Splash \\
NGC~6681 & lowE & 0.923 & secure & low-E & G-2/sg-2 Kraken/Koala \\
NGC~6809 & lowE & 0.908 & secure & low-E & G-8/Pre-disc \\
NGC~6121 & lowE & 0.836 & secure & low-E & G-2/sg-2 Kraken/Koala \\
NGC~6254 & lowE & 0.816 & secure & low-E & G-1/Splash \\
NGC~4833 & lowE & 0.703 & likely & GSE & G-2/sg-5 Pontus \\
NGC~6397 & lowE & 0.621 & likely & M-D & G-1/Splash \\
\end{longtable}
}

\end{appendix}
\end{document}